\documentclass[prb,aps,showpacs,superscriptaddress]{revtex4}
\usepackage{bm}
\usepackage{epsfig}
\usepackage{times}
\usepackage{amssymb}
\usepackage{amsmath}
\usepackage{natbib}
\usepackage{color}
\usepackage[latin1]{inputenc}
\usepackage[loose,nice]{units}       

\begin{document}
\title{Collision and capture dynamics in a quantum dot: The role of doubly excited states}
\author{S{\o}lve Selst{\o}}
\affiliation{Faculty of Technology, Art and Design, Oslo and Akershus University College for Applied Sciences, N-0130
Oslo, Norway}


\begin{abstract}
We investigate collision dynamics involving two electrons and a quantum dot embedded in a quantum wire. One electron is initially at rest in the ground state of the dot, whereas the other electron is incident on the dot with a rather well defined velocity. Excitation, reflection and transmission probabilities are obtained by resolving the dynamics explicitly. Also, the probability of both electrons being unbound after collision is studied.
It is found that the dynamics is strongly influenced by the presence of doubly excited states; reflection becomes considerably more probable in the energetic vicinity of such states. This, in turn, contributes to an increase in the probability of trapping also the second electron within the dot.
\end{abstract}

\maketitle

\section{Introduction}

Since the realization of the first quantum dots about two decades ago, much effort has been spent studying them.
This is motivated by the fact that such devices hold promise for several useful applications. Also, the possibility to {\it design} particular systems with specific features makes the study of such structures an attractive one. Although
these so-called ``artificial atoms'' are similar to atoms in the sense that they share many features, such as, e.g., energy quantization and shell structure, their structure may often be {\it tuned} while the structure of atoms is fixed by fundamental constants.
This opens up a wide area for studying quantum phenomena. Of course, another feature which is appealing to a theorist, is their reduced dimensionality.

A number of phenomena well studied in atoms have later been investigated in quantum dots as well. One such example are doubly excited states embedded in the continuum of a quantum dot. Such states, which are unbound but in some respects resemble bound states, are examples of {\it resonances} states. Their existence have been widely studied, experimentally and theoretically, in all areas where quantum mechanics apply -- also in the context of quantum dots, see, e.g., Refs.~\cite{Pan1994,Gebhard2008,Sajeev2008,Moiseyev2009,Genkin2010}. This is also the aim of the present work. We intend to do so by resolving the explicit time evolution of the system. Although resonances are usually studied in a time independent framework, other dynamical examples are also found in literature, see, e.g. Refs.~\cite{Scrinzi1998,Argenti2010,Bengtsson2012}.

Our system consists of two electron in which one is free and the other is trapped in a quantum dot initially.
As in, e.g., Refs.~\cite{Hausler1993,Gumbs1999,Bednarek2003,Ciftja2006} we will take our system to be confined in two spatial directions so that our problem becomes an effectively one-dimensional one, i.e., the quantum dot is embedded in a quantum wire.
In a collision event like this, there are several possible outcomes. As the system is unbound, at least one electron will escape unless the system is allowed to lose energy somehow. The escaping electron(s) may either be reflected or transmitted. If one electron remains bound, it may be found in an excited state, resulting in an {\it inelastic} scattering event. If the the projectile is incident with sufficient energy, it may liberate also the bound target electron, analogous to {\it ionization} in atomic and molecular physics. Here we will refer to this process as {\it knockout}.

One dimensional scattering problems have been studied extensively -- also with particular attention paid to the role of resonances, see, e.g., Refs.~\cite{Hauge1987,Lindroth2009}. Such studies tend to involve only one particle, i.e., the resonances are {\it shape} resonances.
In the present work we will focus upon how the doubly excited states which in fact are present in our system influence the reflection and transmission probabilities. We will also investigate the capture dynamics in this regard. In Ref.~\cite{Ferreira1999}, which also concerns electron capture and doubly excited states, we may read the following: ``An interesting question is whether and how a second electron can be captured when a first one is already present in the dot.'' This is precisely what we address. In Ref.~\cite{Ferreira1999} two processes are studied: single electron capture by emission of a phonon, and relaxation of one electron via intraband Auger decay, i.e., two electrons in a doubly excited state decay into a state in which one electron is relaxed and the other one escapes.
In this process, no phonon is emitted; the excess energy is carried away by the Auger electron. In the present work we aim to take advantage of both the (intraband) Auger effect {\it and} spontaneous emission in order to capture the projectile electron in the same dot as the target electron.
In fact, this process is the completely analogous to what is called {\it dielectronic recombination} in atomic physics, which, in turn, is the time reverse of Auger decay.

Although our starting point is the Schrödinger equation, the equation we actually solve in the end is the Lindblad equation \cite{Lindblad1976,Gorini1976}. Usually, master equations are used to accommodate for irreversible processes such as, e.g., dissipation. Moreover, in the context of quantum transport, they allow for a convenient description of the coupling between nanostructures and their leads \cite{Harbola2006,Lambert2010}. However, this is not the motivation in the present work. It is simply a consequence of our desire to impose absorbing boundary conditions.

We have chosen to work with units defined by letting $\hbar$, the effective electron mass $m^*$, the elementary charge $e$ and $1/(4 \pi \varepsilon_r \varepsilon_0)$, with $\varepsilon_r$ and $\varepsilon_0$ being the relative permittivity and the permittivity of vacuum, respectively, define the units of their respective quantities. For, e.g., GaAs this corresponds to a length unit of about 10~nm, a time unit of about 60~fs and an energy unit of 11~meV.
These units will be used throughout.

The paper is organized as follows: In Sec.~\ref{theory}, the theoretical framework applied is presented and, in part, derived.
The results are presented and discussed in Sec.~\ref{results}, while conclusions are drawn in Sec.~\ref{conclusions}.

\section{Theoretical framwork}
\label{theory}

The dynamical equations which are to be solved in this work, will deviate considerably from our starting point, which is the Schrödnger equation for two interacting particles in three dimensions. After having reduced the problem to an effectively one-dimensional one, we will, for numerical convenience, introduce absorbing boundary conditions. This calls for the introduction of the Lindblad equation.
Finally, we demonstrate how we estimate the probability of capture through spontaneous emission of a photon or a phonon via the Fermi golden rule.

\subsection{Effective potentials}
\label{EffectivePotentials}

The dynamics is governed by the time dependent Schrödinger equation, which in our units reads
\begin{eqnarray}
\label{SchrodingerV2}
i \frac{d}{dt} \Phi({\bf r}_1,{\bf r}_2;t) & = & \left[ \sum_{i=1}^2 \left(-\frac{1}{2} \, \nabla_i^2 + V({\bf r}_i)\right) + \frac{1}{|{\bf r}_1-{\bf r}_2|} \right] \Phi({\bf r}_1,{\bf r}_2;t) \quad  \text{ with } \\
\label{HiDef3D}
h_i & = & -\frac{1}{2} \, \nabla_i^2 + V({\bf r}_i)  \quad \text{ and} \\
\label{WDef}
W({\bf r}_1, {\bf r}_2) & = & \frac{1}{|{\bf r}_1-{\bf r}_2|} \quad ,
\end{eqnarray}
where $V$ is some local confining potential, i.e. the quantum dot, and $W$ is the electron-electron interaction.

We will restrict the dynamics to one spatial dimension, i.e., the system is confined to the ground states in two Cartesian directions, which we will take to be the $y$ and the $z$ direction.
We will take the confinement to be represented by infinite square wells of equal extension, $l$, in both these directions. Assuming that the confinement is strong enough for us to neglect correlation in these directions, we may, up to a trivial phase factor, write the wave function on the form \cite{Bednarek2003,Ciftja2006}
\begin{eqnarray}
\label{Initial3DWF}
\Phi({\bf r}_1,{\bf r}_2;t) & = & \Psi(x_1,x_2; t)\psi_0(y_1) \psi_0(z_1) \psi_0(y_2) \psi_0(z_2) \quad \text{ with } \\
\nonumber
\psi_0(x) & = & \sqrt{\frac{2}{l}} \cos\left( \frac{\pi}{l} x \right), \quad |x|<l/2 \quad .
\end{eqnarray}
The assumption that the system remains in the ground state in the $y$ and $z$ directions requires that the typical energy transfers involved in the processes we wish to study, $\Delta E$, is considerably smaller than the energy it takes to excite the system in the $y$ or $z$ direction, i.e., $\Delta E \ll 3 \pi^2/(2 l^2)$.

\begin{figure}
  \includegraphics[width=12cm]{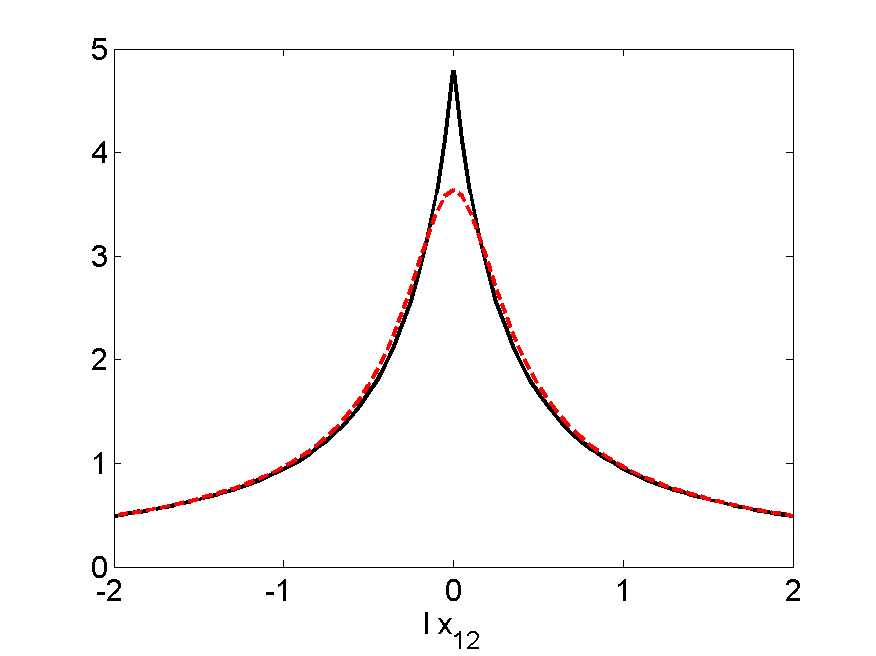}\\
  \caption{(Color online:) Illustration of the effective interaction potential $\widetilde{W}$ (full curve) along with the approximation $\overline{W}$ (dashed curve). The latter has the same asymptotic behavior as the former, and their difference integrates to zero. Specifically, we have here shown $l W(lx_{12})$ (for both functions) such that figure becomes $l$-independent.}\label{InteractionFig}
\end{figure}

The problem is reduced to an effectively one dimensional one by
integrating out the $y$ and $z$ dependencies:
\begin{eqnarray}
\label{SchrodingerV3}
i \frac{d}{dt} \Psi(x_1,x_2; t) & = & \left[ \tilde{h}_1 + \tilde{h}_2 + \widetilde{W}(x_{12})\right] \Psi(x_1,x_2; t)  \quad \text{ with } \\
\label{hiDef1D}
\tilde{h}_i & = & -\frac{1}{2} \frac{d^2}{dx_i^2} +\widetilde{V}(x_i), \quad i\in \{1,2\} \quad , \\
\label{x12Def}
x_{12} & = & |x_1-x_2| \quad , \\
\label{VtildeDef}
\widetilde{V}(x) & = & \int_{-\infty}^\infty dy \int_{-\infty}^\infty dz \, \psi_0(y)^* \psi_0(z)^* \, V({\bf r}) \, \psi_0(y) \psi_0(z) \quad \text{ and} \\
\label{WtildeDef}
\widetilde{W}(x_{12}) & = &
\int_{-\infty}^\infty  dy_1  \int_{-\infty}^\infty dz_1
\int_{-\infty}^\infty  dy_2  \int_{-\infty}^\infty dz_2 \,
\frac{\left|\psi_0(y_1) \psi_0(z_1) \psi_0(y_2) \psi_0(z_2) \right|^2}{\sqrt{x_{12}^2 + (y_1-y_2)^2 + (z_1-z_2)^2}}
\quad .
\end{eqnarray}
The effective interaction may be written as \cite{Ciftja2006}
\begin{equation}
\label{WtildeV2}
\widetilde{W}(x_{12}) =
16 \int_{-1/2}^{1/2} du_1 \int_{-1/2}^{1/2} dv_1 \int_{-1/2}^{1/2} du_2 \int_{-1/2}^{1/2} dv_2 \,
\frac{\cos^2(\pi u_1)\cos^2(\pi v_1)\cos^2(\pi u_2)\cos^2(\pi v_2)}{\sqrt{x_{12}^2 + l^2[(u_1-u_2)^2+(v_1-v_2)^2]}}
\quad ,
\end{equation}
which may be effectively calculated numerically by means of, e.g., Monte Carlo integration. This interaction potential is plotted in Fig.~\ref{InteractionFig} along with the potential
\begin{equation}
\label{SmoothInt}
\overline{W}(x_{12}) = \frac{1}{\sqrt{x_{12}^2+ (l \delta)^2}} \quad .
\end{equation}
In this work, as in several others such as, e.g., Refs.~\cite{Hausler1993,Ciftja2006}, the latter has been applied in the calculations.
This is motivated by the fact that this potential has a continuous derivative, whereas $\widetilde{W}$ features a cusp at $x_{12}=0$, c.f.  Ref.~\cite{Bednarek2003}. The asymptotic behavior is the same for both potentials, and they scale with the confinement $l$ in the same manner \cite{Hausler1993}. The parameter $\delta$ is chosen such that
the difference between $\widetilde{W}$ and $\overline{W}$ integrates to zero.
This is achieved with $\delta \approx$~0.275. We do not see any {\it a priori} reason why replacing $\widetilde{W}$ with $\overline{W}$ should alter the phenomenological features of the system.

As in Ref.~\cite{Sajeev2008}, the effective one-particle potential, $\widetilde{V}$, is chosen to be a negative Gaussian:
\begin{equation}
\label{OnePartPot}
\widetilde{V}(x) = -D_V \exp \left( -\frac{x^2}{\sigma_V^2}\right) \quad ,
\end{equation}
where $D_V$ is the depth and $\sigma_V$ provides the width of the potential.
From a theoretical point of view, this is a convenient choice as it allows for analytical continuation into the complex plane, which will be addressed shortly. Moreover, the short range nature of the potential ensures that there are no Rydberg states.

\subsection{The spectrum}

As pointed out initially, the importance of doubly excited states is crucial to this work. Thus, we need to be able to localize these resonances. This is readily done by the standard technique of complex rotation \cite{Aguilar1971,Balslev1971,Simon1972,Winter1974}; our spatial variables are multiplied by a complex phase factor, $x \rightarrow e^{i \theta}x$, with $0<\theta < \pi/4$. The resulting time independent Schrödinger equation is solved while maintaining Dirichlet boundary conditions.
As is well known, this provides ``eigen energies'' which are unaltered for bound states and rotated the angle $2 \theta$ into the fourth quadrant of the complex plane for ``ordinary'' continuum states. The doubly excited states, however, are identified by the fact that their
complex eigen values are virtually independent of $\theta$ (for large enough $\theta$). The real part of such an ``eigen energy'' gives the position of the resonance, whereas the imaginary part provides its width.

An example of such a complex spectrum is shown in Fig.~\ref{Spektrum}. It corresponds to the system at hand with the parameters $D_V=4$, $\sigma_V=1.5$ and $l=0.7$. The eigen energies of the corresponding one-particle system, i.e., the thresholds, are also displayed.
From this figure we see that the two-particle system features two bound states; the excited state is just barely bound. Morover, one eigen energy with non-vanishing imaginary part is seen to be $\theta$-independent. This complex eigen energy is located at
\begin{equation}
\label{ResonanceEnergy}
E_\mathrm{res} =-1.957 - 0.0896 i \equiv E_\mathrm{res} - i \Gamma_\mathrm{res}/2 \quad \quad (E_\mathrm{res}, \Gamma_\mathrm{res} \in \mathbb{R}) \quad .
\end{equation}

\subsection{Resolving the dynamics}
\label{Dynamics}

\begin{figure}
  \includegraphics[width=12cm]{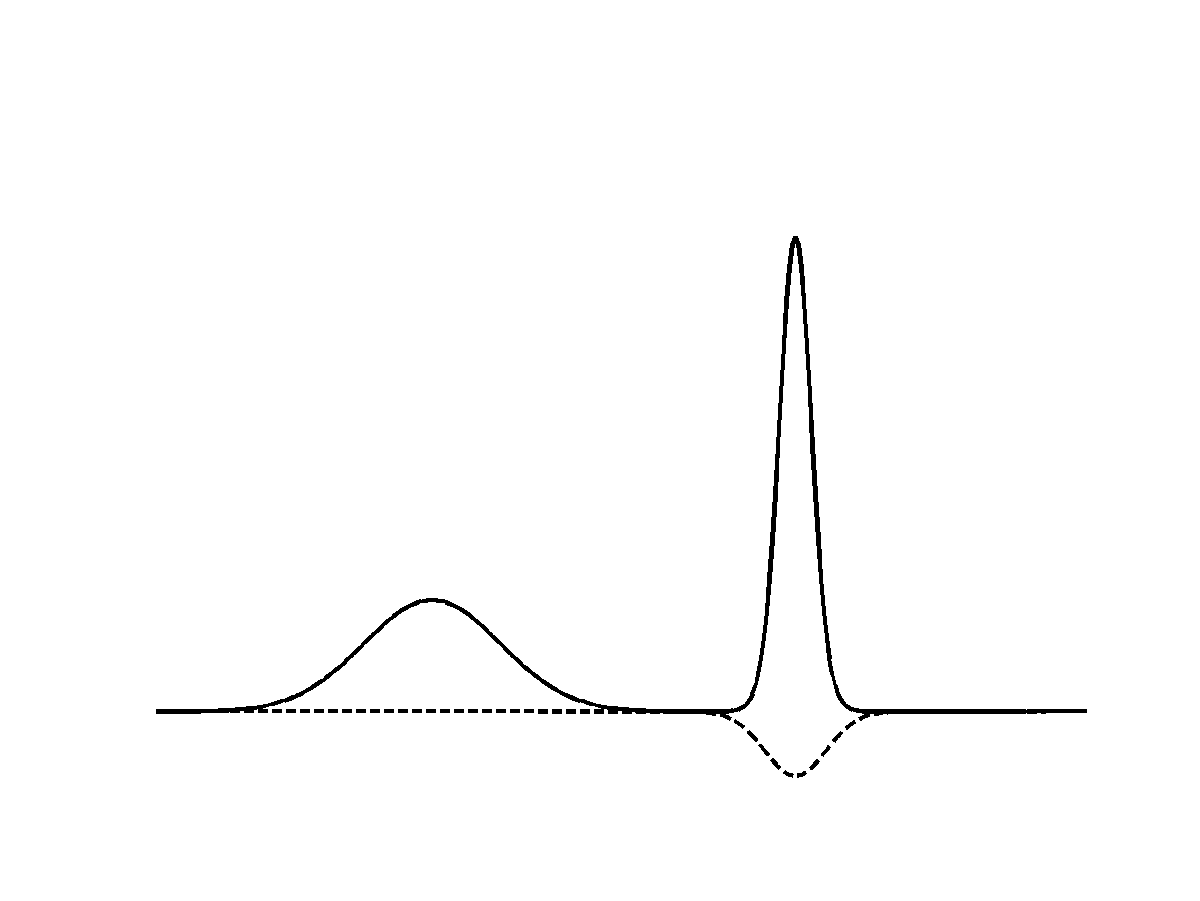}\\
  \caption{The particle density corresponding to our initial state: One electron, represented by a Gaussian, and one electron trapped in a confining potential, which is also a Gaussian (dashed curve).}\label{StartTilstand}
\end{figure}

Our initial state is such that one electron is free and another one is trapped in the (one-particle) ground state for the confining potential $\widetilde{V}$. The free electron is given by a Gaussian wave packet separated from the bound electron. This is illustrated in Fig.~\ref{StartTilstand}.
The packet is centered around a mean velocity $k_0$ with a rather narrow width $\sigma_k$ in momentum space. More specifically, we take the spatial part of our initial state to be
\begin{eqnarray}
\label{ComplexEnergy}
\Psi(x_1,x_2;t=0) & = & \frac{1}{\sqrt{2}} \left[ \phi_{k_0,\sigma_k}(x_1) \, \phi_\mathrm{gs}(x_2) +  \phi_\mathrm{gs}(x_1) \, \phi_{k_0,\sigma_k}(x_2) \right] \quad \text{ where} \\
\label{GaussianWP}
\phi_{k_0,\sigma_k}(x) & = & \frac{1}{\pi} \, \sqrt{\frac{\sigma_k}{1-i \sigma_k^2t_0}} \, \exp \left[ -\frac{\sigma_k^2 (x-x_0)^2}{2(1-i \sigma_k^2t_0)}+ i k_0(x-x_0)\right] \quad  ,
\end{eqnarray}
$x_0$ is the initial centering of the incoming electron, and $t_0$ is the instant a freely propagating wave packet is at its narrowest (spatially).
$\phi_\mathrm{gs}(x)$ is the one-particle ground state. As seen from the exchange symmetry of the above state, we have chosen our initial state to be a spin singlet state.
The mean energy of the incoming particle is given by
\begin{equation}
\label{MeanE}
\overline{E} = \quad \frac{k_0^2}{2} + \frac{\sigma_k^2}{4} \approx \frac{k_0^2}{2}
\end{equation}
with standard deviation provided by
\begin{equation}
\label{DeviationE}
\sigma_E^2 = \frac{\sigma_k^2}{4} \left( 2 k_0^2 + \frac{\sigma_k^2}{2} \right)  \approx \frac{\left(k_0 \sigma_k \right)^2}{2}\quad .
\end{equation}
The total energy of the system is
\begin{equation}
\label{EnergySyst}
E_i=\overline{E} + \varepsilon_\mathrm{gs} \quad ,
\end{equation}
where the last term is the one-particle ground state energy.

Since the incoming Gaussian is rather narrow in momentum space, it must be quite wide in position space. Thus, a numerical grid of rather large extension is needed. During propagation, this situation is complicated further due to dispersion and the fact that parts of the wave packet will be reflected. In terms of feasibility, this may be problematic for any system involving more than one particle -- even in the one dimensional case.

Thus it is tempting to introduce absorbing boundary conditions as a numerical tool. However, doing so introduces other problems: When absorption takes place, the entire wave function, which is normalized to the probability of having two particles on the grid, vanishes. If we want to know what happened to the possibly remaining electron, we are at a loss. We will, e.g., not be able to know whether one or two electrons have escaped, or what the excitation probabilities are. As we also aim to calculate the probabilities of such events, we are indeed faced with this problem.

As it turns out, the solution to this problem is provided by the Lindblad equation \cite{Lindblad1976,Gorini1976}. This is explained in detail in Ref.~\cite{Selsto2010}. Here we will very briefly outline the main ideas. The Lindblad equation is the natural starting point as it ensures conservation of positivity and trace for a Markovian process, such as absorption. By comparing the effective Hamiltonian involving a complex absorbing potential with the generic structure of the Lindblad equation, a source term is identified, which ``feeds'' a one-particle density matrix $\rho_1$ as the two particle state vanishes due to absorption. The one-particle system, in turn, is also exposed to the absorber, which may lead to population of the vacuum-state $\rho_0=p_0|-\rangle \langle - |$, i.e., the state in which there are no particles. The total density matrix features ``super selection'', i.e., if the initial state corresponds to a well defined number of particles, only ``diagonal blocks'' are populated;
\begin{equation}
\label{BlockStructure}
\rho = \sum_{n=0}^2 \rho_{n,n} \quad ,
\end{equation}
where $n$ corresponds to the number of particles.

\begin{figure}
  \centering
  \includegraphics[width=12cm]{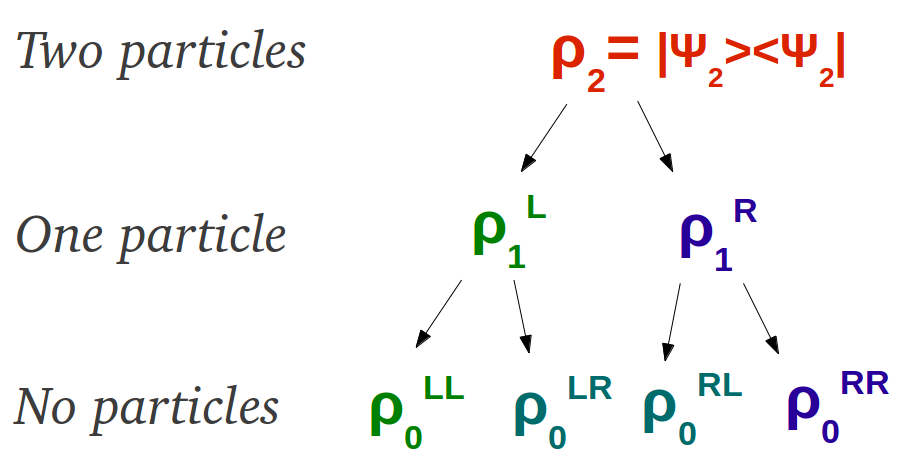}\\
  \caption{(Color online:) Schematic illustration of the various ways of reducing the number of particles. We start out with two particles, which may be turned into a one-particle system by absorption either to the left or the right. The remaining electron may also, in turn, be absorbed, thus populating the vacuum state.}\label{ReduksjonsSkjema}
\end{figure}

In this work, we will modify the theory of Ref.~\cite{Selsto2010} slightly in the sense that we will discriminate between the left and right absorber. We will distinguish the one-particle density matrix in a part arising from absorption at the left edge and one arising from absorption at the right edge,
\begin{equation}
\label{SplitRho1}
\rho_1=\rho_1^\mathrm{L}+\rho_2^\mathrm{R} \, .
\end{equation}
Similarely, we may distinguish the population of the vacuum state, $p_0$, in four parts corresponding to the four ways of absorbing both electrons. This is illustrated in Fig.~\ref{ReduksjonsSkjema}.
Specifically, in our grid representation, the dynamical equation may be written as
\begin{eqnarray}
\label{LindbladEquation}
i \dot{\rho} & = & \left[ H, \rho \right] - i\mathcal{L}_\mathrm{L}[\rho] - i\mathcal{L}_\mathrm{R}[\rho] \quad \text{ with the Lindbladians} \\
\nonumber
\mathcal{L}_\mathrm{L/R} & \equiv & \sum_n \Gamma_\mathrm{L/R}(x_n) \left( \left\{c_n^\dagger c_n, \rho \right\} - 2 c_n \rho c_n^\dagger \right) \quad ,
\end{eqnarray}
where $c_n^{(\dagger)}$ annihilates (creates) a particle in position $x_n$ and $\Gamma_\mathrm{L/R}$ is the left/right absorber. $H$ is the total Hamiltonian expressed in terms of second quantization.

In this work we have chosen to use an absorber of the form
\begin{equation}
\label{absorber}
\Gamma_\mathrm{R}(x) = \left\{ \begin{array}{ll} 2.5 k_0^2 \, \frac{(x- L/2 +l_\mathrm{abs})^2}{l_\mathrm{abs}^2},  & L/2-l_\mathrm{abs}< x <L/2
\\ 0, & \text{otherwise} \end{array} \right. \quad ,
\end{equation}
where $l_\mathrm{abs}=L/10$, i.e., the absorber is `turned on' at distances one tenth of the extension of the grid from the edges, and it increases quadratically towards $2.5 k_0^2 \approx 5 \overline{E}$ at the edge right edge at $x= L/2$. The left absorber is defined correspondingly, and their sum constitutes the total absorber, $\Gamma=\Gamma_\mathrm{L}+\Gamma_\mathrm{R}$. We have chosen the strength of the absorber proportional to the mean energy of the incoming electron in order to achieve sufficient absorption while inducing minimal unphysical reflection.

If we write down the evolution of the various constituents of the total density matrix separately, it is seen that the two-particle part remains a pure state, $\rho_2 = | \Psi_2 \rangle \langle \Psi_2|$ where the wave function obeys the ordinary non-Hermitian Schrödinger equation,
\begin{equation}
\label{TwoPartPart}
i \dot{\Psi}_2 = \left[\tilde{h}_1 + \tilde{h}_2 + \widetilde{W}(x_{12})- i\Gamma(x_1) - i\Gamma(x_2)\right] \Psi_2 \quad .
\end{equation}
The one-particle parts are given by
\begin{equation}
\label{OnePartPart}
i \dot{\rho}_1^D = [\tilde{h}, \rho_1^D] - i \{ \Gamma, \rho_1^D \} + 2i \sum_n \Gamma_D(x_n) \, c_n |\Psi_2 \rangle \langle \Psi_2 | c_n^\dagger
\end{equation}
and the vacuum probability is given by
\begin{equation}
\label{ZeroPartPart}
\dot{p}_0^{DE}  = 2 \langle - | \sum_n \Gamma_E(x_n)  c_n \rho_1^D c_n^\dagger| - \rangle  \quad .
\end{equation}
The indices `$D$ and $E$' stands for either `left' or `right'($\mathrm{L}/\mathrm{R}$).

\subsection{Decay by spontaneous emission}

This work aims to investigate whether doubly excited states may be used actively in trapping electrons. Of course, as long as the dynamics does not involve any interaction with any field, energy is conserved and capture is prohibited. Thus, the possibility of relaxation via emission of some field quanta must be included in our calculations somehow.
Of course, this may be achieved by actually including some quantized field in our formalism. This is numerically infeasible as this would introduce too many additional degrees of freedom in our dynamical calculations.
Another way would be to describe the field classically, i.e., non-quantized, which is admissible if a large number of field quanta are present initially. However, we are interested in {\it spontaneous} emission in this context.

With our dynamics already described by an equation of Lindblad form, an approach in which the degrees of freedom of the quantized field is traced out, resulting in a Master equation for the reduced density matrix, could be a fruitful path \cite{Mollow1975,Plenio1998}. However, since our two-particle wave function has a rather well defined energy, we may resort to a simpler strategy: We will estimate dynamical capture rates based on the Fermi golden rule. In the limit that the electronic system couples weakly to the field, the rate at which the system is captured into bound state $b$ by spontaneous emission of a field quantum is estimated by
\begin{equation}
\label{FermiGoldenRule}
\dot{P}_b(t) = 2 \pi \sum_{\bf k} \left| \langle \Phi_b, 1_{\bf k} | H_I | \Psi_2(t), 0 \rangle \right|^2 \,
\delta\left[ k_0^2/2 + \varepsilon_0^{(1)} - (\varepsilon_b^{2} + \hbar \omega(k)) \right] \quad ,
\end{equation}
where the sum runs over all modes ${\bf k}$. $|\Phi_b, 1_{\bf k} \rangle$ is the composite state in which the electronic part is in bound state $b$ and one quantum has been emitted in mode ${\bf k}$, $| \Psi_2(t), 0 \rangle$ corresponds to the electronic state $\Psi_2$ with zero field quanta. Thus, by calculating the time integral of Eq.~(\ref{FermiGoldenRule}), the probability of capture through spontaneous phonon emission may be estimated.

We should also address the interaction $H_I$. The interaction with the phonon field may be represented by
\begin{equation}
\label{LengthGaugeInteraction}
H_I^\mathrm{photon} = i {\bf r} \cdot \sum_{{\bf k},\mu} \sqrt{\frac{2 \pi}{\omega(k) V}}
\left( {\bf e}_{\bf{k}, \mu} \hat{a}_{\bf{k}, \mu} e^{i {\bf k}\cdot {\bf r}} -
{\bf e}_{\bf{k}, \mu} \hat{a}_{\bf{k},\mu}^\dagger e^{-i {\bf k}\cdot{\bf r}}\right) \quad ,
\end{equation}
where $V$ is the quantization volume and $\hat{a}_{\bf{k}, \mu}^{(\dagger)}$ annihilates (creates) a photon in the mode given by the wave vector ${\bf k}$ and polarization $\mu$. However, in a solid state context, we would usually expect the the interaction with phonons to be more important. In the case of the deformation potential, this interaction reads
\begin{equation}
\label{DeformationPotential}
H_I^\mathrm{phonon} = \lambda \sum_{\bf k} \frac{k}{\sqrt{\omega(k)}} \left( \hat{a}_{\bf{k}} e^{i {\bf k}\cdot {\bf r}} + \hat{a}_{\bf{k}}^\dagger e^{-i {\bf k}\cdot{\bf r}}\right) \quad ,
\end{equation}
where the parameter $\lambda$ depends on the volume of the sample, the density of the material and the strength of the deformation potential tensor \cite{Seeger1997}.

Although seemingly rather similar in form, these interactions have quite different features. In the case of the photon interaction, the wave number in the exponential is related to the energy shift $\omega$ by $k = \omega/c$ with $c$ being the speed of light. As this is a large number and $\omega$ is typically moderate, we may safely approximate the exponential factors in Eq.~(\ref{LengthGaugeInteraction}) by unity, thus removing the explicit energy dependence of the interaction. This approximation, i.e., the {\it dipole approximation}, is one that only couples states of opposite parity.
When it comes to the phonon interaction, Eq.~(\ref{DeformationPotential}), the energy shift $\omega$ and the wave number $k$ is related by dispersion relations, which are specific to the material at hand and the the kind of phonons. On our length scale, typical $k$-values may become quite large, thus resulting in a very strong energy dependence.

\subsection{Numerical implementation}

The two-particle wave function is calculated using a rather standard spilt operator technique \cite{Feit1982}. We have split the propagator according to the number of particles it acts on:
\begin{eqnarray}
\label{TwoParticlePropagator}
\Psi_2(t+\tau) & = &
e^{-iA \tau/2} \, \exp(-i \overline{W}(x_{12}) \tau) \, e^{-iA \tau/2} \, \Psi_2(t) +\mathcal{O}(\tau^3)
\quad \text{ with}\\
\nonumber
A & \equiv & \tilde{h}_1+ \tilde{h}_2 - i [\Gamma(x_1)-\Gamma(x_2)] \quad .
\end{eqnarray}
We represent the wave function as a symmetric matrix with indices given by the position of each particle.
In this way, the two-particle operator is implemented as entry-wise multiplication and the one-particle operators are implemented as left and right multiplication, respectively, with the state matrix.

In earlier implementations of the Lindblad equation for an initial two-particle system with absorbing boundaries, the numerical scheme has been suffering from not being manifestly trace-conserving \cite{Selsto2010,Selsto2011}. Thus, as conservation of probability is crucial, it was necessary to resort to a very small time step $\tau$. In the present work this problem has been considerably reduced. We use the following scheme, based on a three point formula for integration, to solve Eq.~(\ref{OnePartPart}):
\begin{eqnarray}
\rho_1^D(t+2 \tau) & = & e^{-2 i B \tau} \rho_1^D(t) e^{2 i B^\dagger \tau} + \\
\nonumber
& & \frac{\tau}{3} \left(e^{-2 i B \tau}S_D(t)e^{2 i B^\dagger \tau} + 4e^{- i B \tau}S_D(t+\tau)e^{i B^\dagger \tau} + S_D(t+2\tau) \right)+\mathcal{O}(\tau^5) \quad \text{ with}\\
\nonumber
B & \equiv & \tilde{h}-i \Gamma \quad \text{ and} \\
\nonumber
S_D(t) & \equiv & 2 \sum_n \Gamma_D(x_n) c_n | \Psi_2(t) \rangle \langle \Psi_2(t)| c_n^\dagger \quad .
\end{eqnarray}
As before, ``$D$'' is either L for `left' or R for `right'.

The four parts of the zero-particle probability, $p_0^{DE}$, are simply obtained by integrating Eq.~(\ref{ZeroPartPart}). Also this may be done by means of the three point formula. Examining to what extent the total trace, $\mathrm{Tr} \rho = |\Psi_2|^2 + \mathrm{Tr} \rho_1 + p_0$ remains unity at all times provides a good check for the accuracy of our calculations.

We will need the (non-complex scaled) wave functions of all bound states in order to be able to calculate capture rates, c.f. Eq.~(\ref{FermiGoldenRule}). These states are, along with their energies, determined by propagation in imaginary time, i.e., by replacing $\tau$ by $-i \tau$ in Eq.~(\ref{TwoParticlePropagator}).
With this substitution, an arbitrary initial state with some symmetry property will exponentially converge towards the lowest state within that symmetry, and its energy is provided by the change in norm at each time step. Within a specific symmetry, excited states are found by repeatedly projecting away already obtained lower lying states.

\subsection{Analysing the state}

 For large times $t$, the traces of each of the constituent parts of the density matrix are subject to rather straight forward interpretation. The trace of $\rho_1^\mathrm{L}$, e.g., converges towards the probability that one, and only one, electron has been absorbed and that it has been absorbed to the left. I.e., it converges towards the reflection probability $R$. Similarly, $\mathrm{Tr} \, \rho_1^\mathrm{R}$ provides the transmission probability $T$:
\begin{eqnarray*}
\label{ReflectionCoefficient}
R & = & \displaystyle\lim_{t\to \infty} \mathrm{Tr} \, \rho_1^\mathrm{L}(t) \\
\label{TransmissionCoefficient}
T & = & \displaystyle\lim_{t\to \infty} \mathrm{Tr} \, \rho_1^\mathrm{R}(t) \quad .
\end{eqnarray*}
With an unbound initial state and no field interaction the norm of the two-particle wave function will necessarily converge towards zero. Moreover, if knockout is energetically prohibited, the reflection and transmission coefficients may be found without calculating $\rho_1$ explicitly. To see this, consider the evolution of the trace as dictated by Eq.~(\ref{OnePartPart}):
\begin{equation}
\label{TraceNoRho1}
\mathrm{Tr} \, \dot{\rho}_1^D =
2 \langle \Psi_2 | \left[ \Gamma_D(x_1) + \Gamma_D(x_2)\right] | \Psi_2 \rangle -
2 \mathrm{Tr} \left[ \Gamma \rho_1^D \right] \quad .
\end{equation}
If we know that $\rho_1$ never overlaps with the absorber, the latter term vanishes and we may find $T$ and $R$ simply by integrating the overlap between the two-particle wave function and the respective parts of the absorber as functions of time,
\begin{equation}
\label{RTcoeffNoRho1}
R \approx 2 \int_0^\infty \langle \Psi_2 (t) | [\Gamma_\mathrm{L}(x_1)+\Gamma_\mathrm{L}(x_2) ] | \Psi_2(t) \rangle \, dt \quad ,
\end{equation}
with $T$ calculated analogously.

Although this rather intuitive relation may be useful, a lot more information may be obtained in the framework provided by the Lindblad equation. We may, e.g., find the probability of exciting the electron that remains after collision and the probability that both electrons are unbound after collision. The latter is simply provided by $p_0$ in the limit $t \rightarrow \infty$, whereas the population of bound one-particle state number $b$ is found as
\begin{equation}
\label{Pexcite}
P_\mathrm{excite}^{(b)}(t) =  \mathrm{Tr} \, \left[ |\varphi_b \rangle \langle \varphi_b | \rho_1(t) \right] = \langle \varphi_b | \rho_1(t) | \varphi_b \rangle \quad .
\end{equation}
As we have seen, all of these quantities may also be found differentially in terms of the direction(s) of the escaping electron(s). The knockout probability may also be found in a manner which converges faster in time:
\begin{equation}
\label{DoubleKnockoutV2}
P_\mathrm{ko} =
\displaystyle{\lim_{t \rightarrow \infty}} \left( p_0(t) + \mathrm{Tr}[\mathcal{P}_\mathrm{UB} \rho_1(t)] \right) \quad ,
\end{equation}
where $\mathcal{P}_\mathrm{UB}$ projects onto the subspace of unbound one-particle states. The latter formula does not, however, provide any information about the direction in which the second electron is ejected.

As the absorbers remove all information about escaping particles, the system is bound to lose coherence. This loss may be quantified, e.g., by the von Neumann entropy, $S=-\mathrm{Tr} \, [\rho \log_2 \rho]$, or by the purity, $\zeta = \mathrm{Tr} \, \rho^2$, which is unity for pure states and decreasing with increasing ``mixedness''. Since $\rho$ is block-diagonal in term of particle number, c.f. Eq.~(\ref{BlockStructure}), this quantity may be calculated as the sum of the purity of each block:
\begin{equation}
\label{Purity}
\zeta = |\Psi_2|^4 + \mathrm{Tr} \, \rho_1^2 + p_0^2 \quad .
\end{equation}

\section{Results and discussion}
\label{results}

\begin{figure}
  \includegraphics[width=14cm]{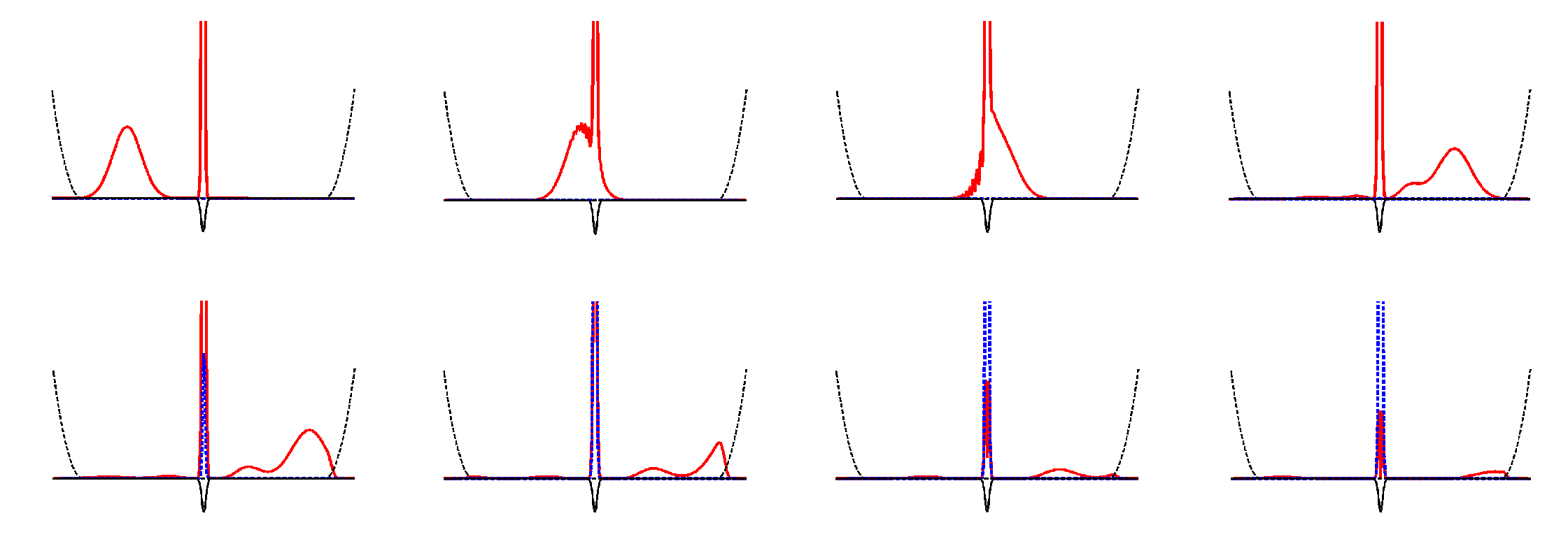}\\
  \caption{(Color online:) The particle density at various times during the collision process. The full (red) curve is the particle density originating from the two particle wave function, whereas the dashed (blue) curve is the particle density corresponding to the one-particle sub-system. Also illustrated are the complex absorbing potential and the confining potential (thin curves).}\label{Cartoon}
\end{figure}

In order to illustrate the collision dynamics, a series of ``snapshots'' of the particle density is displayed in Fig.~\ref{Cartoon}.
The full curve is the particle density given by the two particle wave function, i.e. $\int_{-\infty}^\infty |\Psi_2(x_1,x;t)|^2 \, dx_1$, and the dashed curve is the particle density of the one-particle sub-system. The latter is simply the diagonal of $\rho_1$. In the absorption free region of the grid, the sum of these densities coincide with the particle density obtained from a wave function obtained on an untruncated grid. The incoming electron has a mean energy $\overline{E}=1.8$ with $\sigma_k=0.1$.
In these calculations, a grid extending over $L=150$ length units have been used, featuring 511 grid points. Converged results were obtained with the numerical time step $\tau = 0.05$.

\begin{figure}
  \includegraphics[width=12cm]{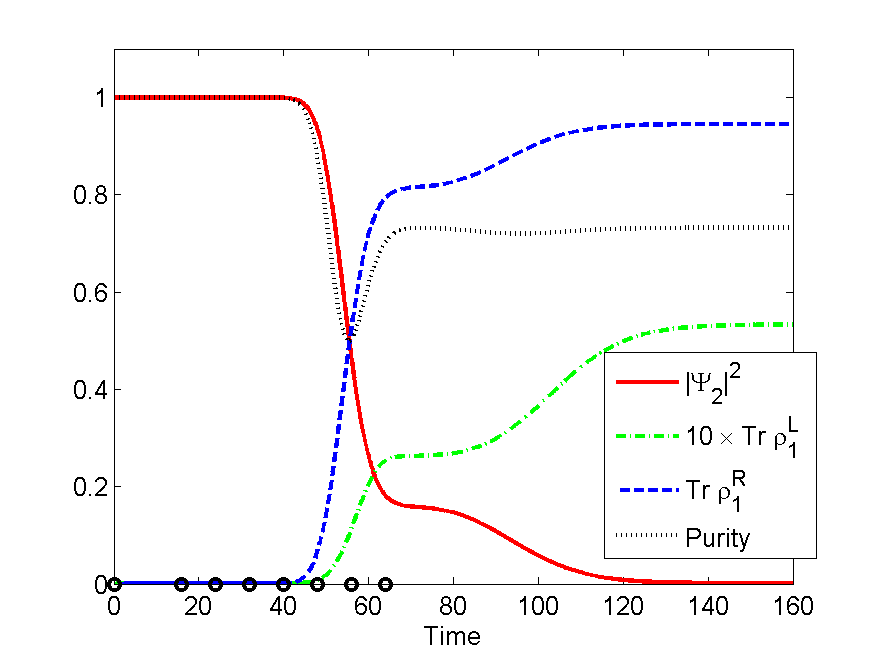}\\
  \caption{(Color online:) These curves refer to the collision process depicted in the ``cartoon'' in Fig.~\ref{Cartoon}. The full curve shows the norm of the two-particle wave function, the dashed one is the trace of $\rho_1^\mathrm{R}$, i.e. the probability of right absorption, the dash-dotted one is $\mathrm{Tr} \, \rho_1^\mathrm{L}$ multiplied by 10. The purity of total density matrix is shown as a dotted curve. The instants corresponding to the various ``snapshots'' in Fig.~\ref{Cartoon} are indicated by circles.}\label{Norms}
\end{figure}

As is seen, the one-particle sub-system is populated as the two-particle wave function is absorbed. As the system is completely unbound, $\Psi_2$ eventually vanishes completely, leaving all population in $\rho_1$ (knockout is not energetically accessible in this case). This is also illustrated in Fig.~\ref{Norms}, which shows the norm of the two-particle wave function as a function of time -- along with the traces of $\rho_1^\mathrm{L}$ and $\rho_1^\mathrm{R}$. The instances shown in Fig.~\ref{Cartoon} are indicated at the $x$-axis. We have also included the evolution of the purity, c.f. Eq.~(\ref{Purity}), of the system.
As this quantity converges towards a value below unity, it is clear that coherence is lost in the process, and the final state is not a pure state.

In Fig.~\ref{Norms} we observe that there is a certain probability of reflection in this collision. Although just barely, this may be observed in Fig.~\ref{Cartoon} as well. In this specific case, the reflection probability is 5.3\%. In Fig.~\ref{Cartoon} it is seen that the escaping electron comes out in two ``lobes'' -- corresponding to different group velocities. In Fig.~\ref{Norms} this is manifested in a step-like increase in the two one-particle traces. This is due to the fact that the incident electron has enough energy to excite the bound one. For the slow lobe, this excitation has taken place, whereas the faster lobe corresponds to elastic scattering.

\begin{figure}
  \includegraphics[width=12cm]{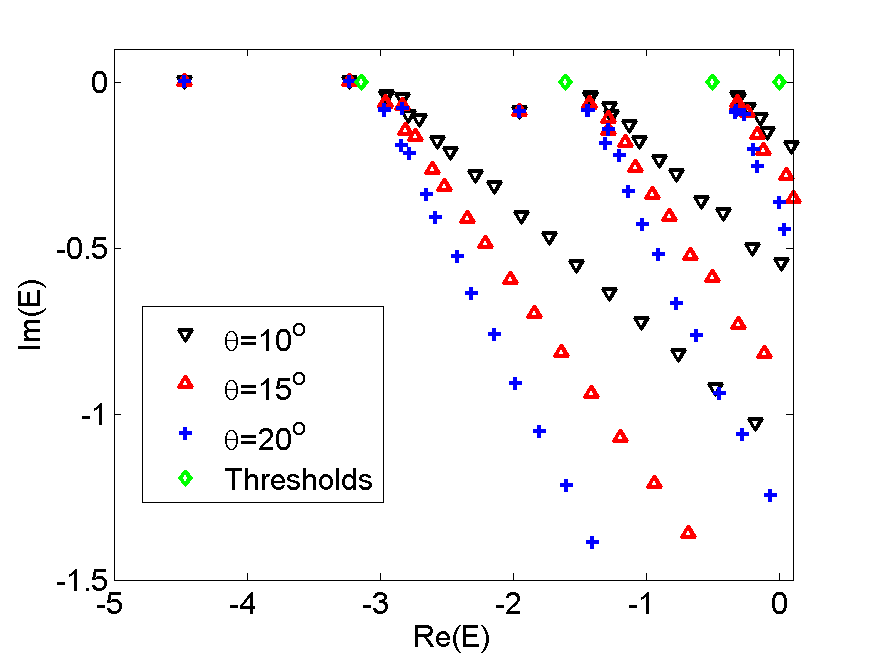}\\
  \caption{(Color online:) The spectrum of the complex rotated Hamiltonian for three values of $\theta$. The eigen energies of the corresponding one-particle system, i.e. the thresholds, are also displayed.}\label{Spektrum}
\end{figure}

\begin{figure}
  \includegraphics[width=12cm]{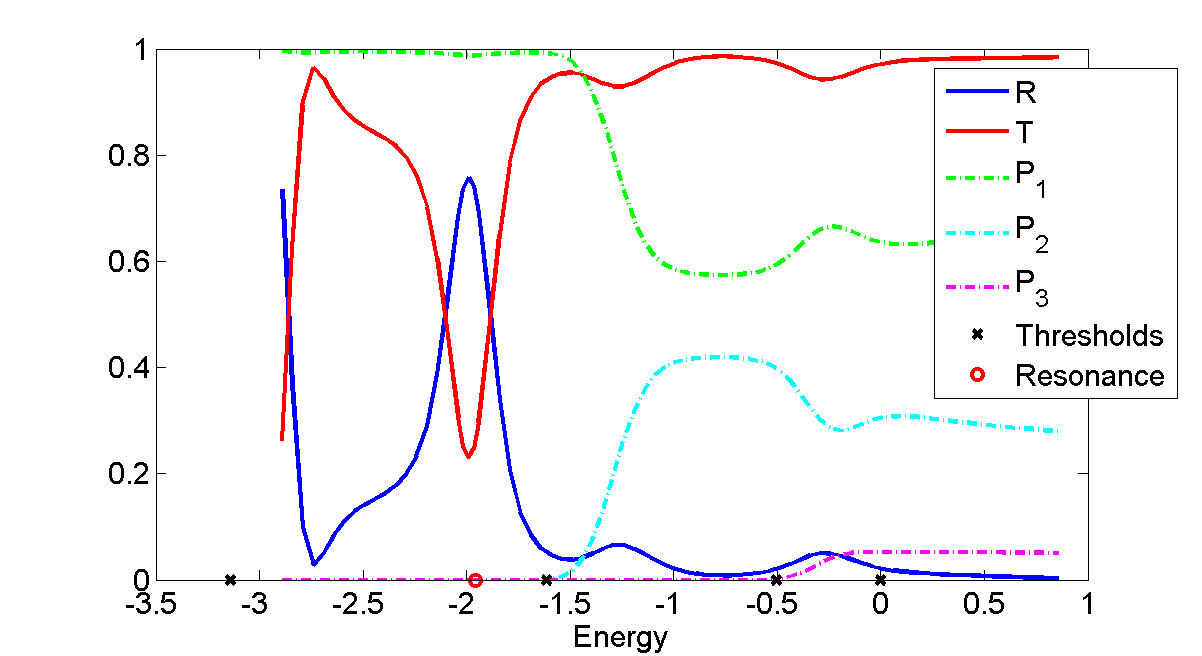}\\
  \caption{(Color online:) The transmission and reflection coefficients, $T$ and $R$, respectively, as functions of the energy of the initial two-particle wave function, $E_i=k_0^2/2 + \varepsilon_\mathrm{gs}$ (full curves). The dash-dotted curves show the probability for the system to end up in the various one-particle bound states. The x-marks on the $x$-axis indicate the onset of each of these channels -- except for the first one, which is the one-particle ground state energy. The red circle is the real part of the ``eigen energy'' og the resonance.}\label{MainResult}
\end{figure}

In Fig.~\ref{MainResult} we have shown how the reflection and transmission coefficients depend on the the energy $E_i$
of the two-electron system. Here the one-particle ground state energy is $\varepsilon_\mathrm{gs} = -3.141$. We have also displayed the probability for the system to end up in the various one-particle bound states. The parameters of the potential $\widetilde{V}$ and interaction $\widetilde{W}$ are the same ones as those of Fig.~\ref{Spektrum}. The incident electron has $\sigma_k=0.06$, c.f. Eq.~(\ref{GaussianWP}). Converged results where obtained using a grid consisting of 1023 points extending over $L=250$ length units. As in Fig.~\ref{Cartoon}, we have used the numerical time step $\tau=0.05$.

It is seen that for electrons incident with energies below the energy required for excitation, only elastic scattering takes place, i.e. the one-particle ground state is the only populated state after collision. When the inelastic channels open, the populations of the excited states become appreciable. The dependence on initial energy is not monotonic, however. Their structure seems to be related to the onset of each channel, whereas the populations of the various bound states feature more or less plateau-like behavior in between.

\begin{figure}
  \includegraphics[width=12cm]{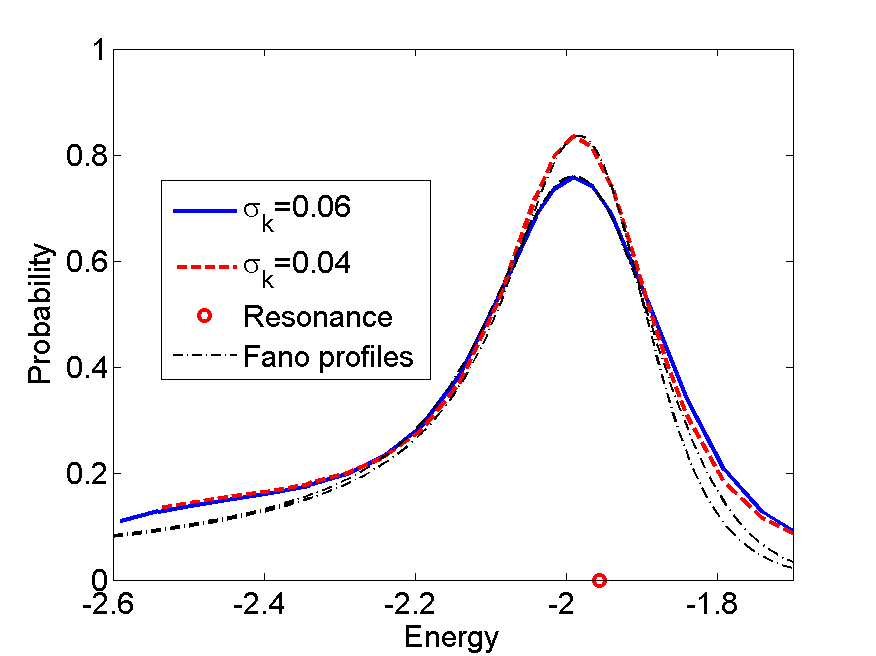}\\
  \caption{(Color online:) A close up on the peak seen in $R(E_i)$ in Fig.~\ref{MainResult} (full curve) along with a convoluted Fano-profile (dash-dotted curve). Also included are the corresponding curves for a case in which the Gaussian wave packet of the projectile electron has a narrower momentum distribution ($\sigma_k=0.04$). As in Fig.~\ref{MainResult}, the circle on the $x$-axis marks the position of the resonance state.}\label{FanoFits}
\end{figure}

The most striking feature of Fig.~\ref{MainResult}, however, is the pronounced peak in the reflection coefficient $R(E_i)$, or, correspondingly, the dip in $T$, at $E_i \approx E_\mathrm{res}$. This is a clear manifestation of the doubly excited state seen in Fig.~\ref{Spektrum}.
A careful look, however, reveals that the maximum in $R$ does not completely coincide with $E_\mathrm{res}$; it is shifted slightly to the left. As is indicated in Fig.~\ref{FanoFits}, this may be due to the interference between the resonance and the background continuum.
Here we have plotted a close up of the peak together with a convoluted Fano profile \cite{Fano1961}. The top of the peaks coincide for a $q$-value of $-4.5$. This parameter and the hight of the peak are the only two parameters used in the fitting. The rest is provided by the energy-width of the incident electron, $\sigma_k$, and the resonance parameters $E_\mathrm{res}$ and $\Gamma_\mathrm{res}$, c.f. Eqs.~(\ref{GaussianWP},\ref{ResonanceEnergy}). Performing a convolution is necessary as the width in the energy distribution of the incident electron is comparable to the width of the resonance, $\Gamma_\mathrm{res}$. A second case with a narrower energy distribution, $\sigma_k=0.04$, is also included.

In the expression for the initial Gaussian wave packet, Eq.~(\ref{GaussianWP}), the parameter $t_0$ is the time at which the packet, if allowed to propagate freely, is at its narrowest. In all the above cases, this has been set to zero, i.e., the projectile wave packet widens as it approaches the target. One may ask to what extent the collision dynamics are altered if we chose a different $t_0$. We have repeated the above calculations with a $t_0$ which is such that the projectile is at its narrowest at impact. For the quantities presented in Figs.~\ref{MainResult} and \ref{FanoFits} no differences where seen.

\begin{figure}
  \includegraphics[width=12cm]{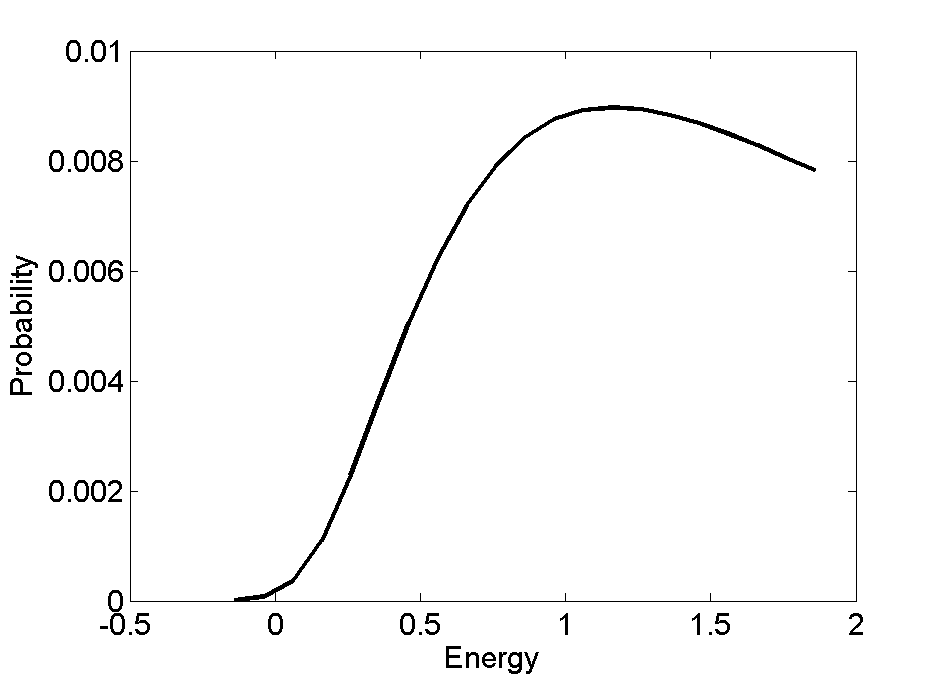}\\
  \caption{The probability of both electrons being liberated as a function of the energy of the system.}\label{DKO}
\end{figure}

For projectile energies larger than the ionization potential of the target electron, knockout becomes energetically possible. In Fig.~\ref{DKO} the probability of this process is shown as a function of the initial energy for the same case as in Fig.~\ref{MainResult}. A finite probability for knockout is seen for energies beyond zero. It does not increase monotonously with energy. It is also seen that the probability stays rather low. For this reason a lower numerical time step had to be used than in the case of Fig.~\ref{MainResult}. The low probability of knockout may be understood from the fact that numerical solutions of Newton's second law for the system suggest that the process does not happen at all classically. When the process {\it does} have a finite probability quantum mechanically, it may be related to the fact that classically, the projectile electron may gain energy on the expense of the target electron, whereas this is impossible in a quantum mechanical context as the target electron is in its ground state initially.

The fact that reflection becomes strongly enhanced due to the presence of doubly excited states suggest that also the capture probability may be increased by such resonances. Since a projectile electron which eventually is reflected typically ``lingers'' around the target electron and the confining potential longer than a transmitted one, there should be more time for the free electron to ``fall into'' the confining potential by spontaneous emission of a phonon or a photon.

\begin{figure}
  \includegraphics[width=12cm]{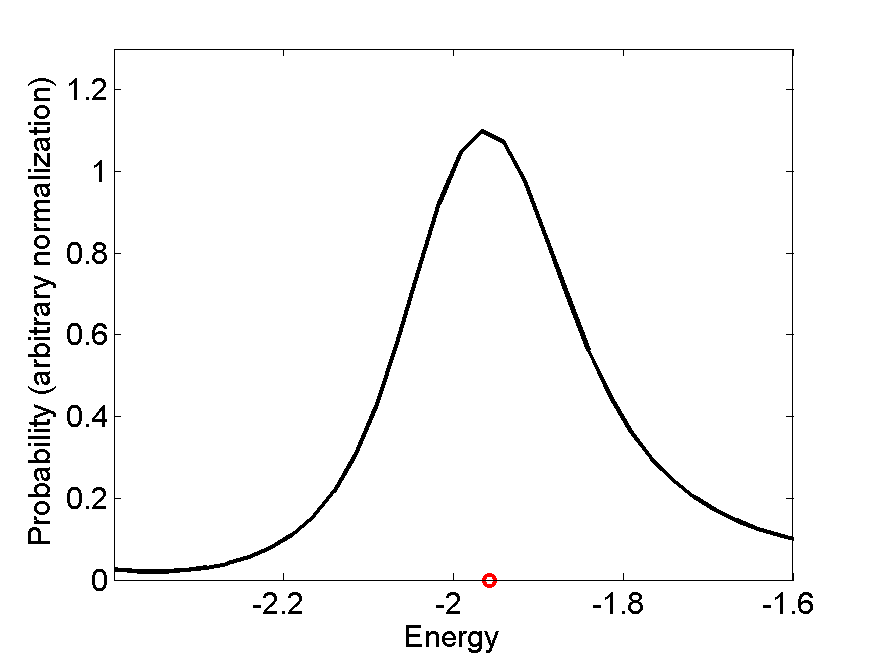}\\
  \caption{Estimate of how the probability of capture via spontaneous emission of a photon behaves in the vicinity of the doubly excited state. The position of the doubly excited state is indicated by a circle.}\label{CaptureCaseOne}
\end{figure}

Figure~\ref{CaptureCaseOne} shows how the probability of capture due to spontaneous emission of a photon depends on energy of the initial state. As with the reflection coefficient, a clear peak is seen in the vicinity of the resonance state. Also in this case, a shift towards lower energies as compared to $E_\mathrm{res}$ is seen. This may, once again, be due to the interference, but it could also be due to the fact that as the ``overlap time'' in general decreases as the group velocity of the projectile electron increases. For this reason, the capture probability is an over all decreasing function of energy -- except in the vicinity of resonances.

Also for the phonon interaction, the capture probability is a decreasing function of energy. However, here the effect mentioned above is completely dominated by another one. Due to the before mentioned strong explicit energy dependence in the interaction Eq.~(\ref{DeformationPotential}), the probability of relaxation via phonon emission falls off quite dramatically. For this reason, no resonance peak is observed for phonons for the case at hand. We do, however, still expect to see such peaks at smaller energy scales.

\begin{figure}
  \includegraphics[width=9cm]{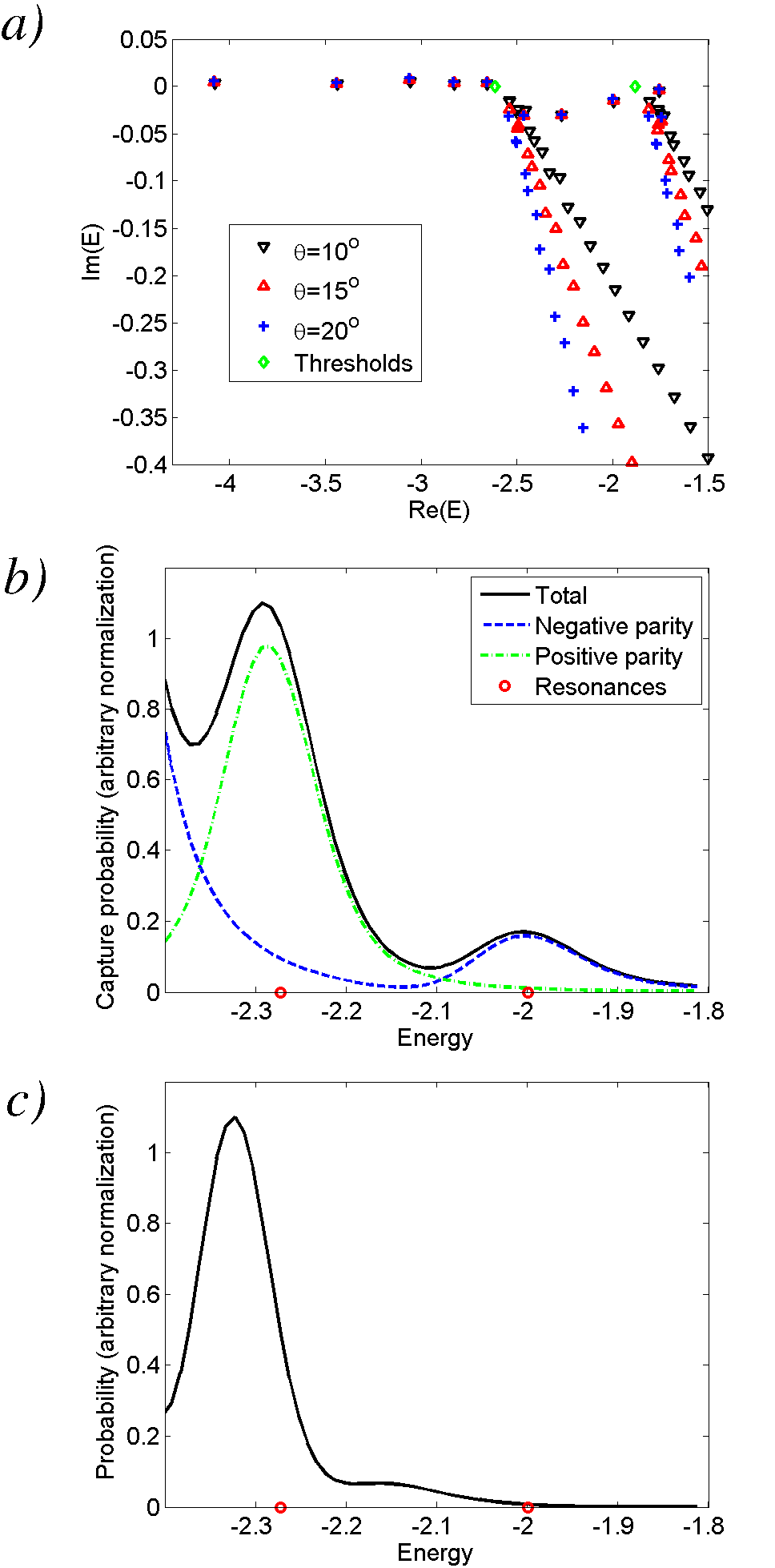}\\
  \caption{(Color online:) {\it a)} The complex spectrum of the case at hand for three different scaling angles. Two resonance states are identified between the first and second threshold. {\it b)} Estimates of the probability of capture via photon emission. The partial contributions from bound states of positive and negative parity are also displayed. {\it c)} Estimates of the probability of captured via phonon emission. In the two lower panels, the positions of the two resonances are indicated.}\label{CaptureCaseTwo}
\end{figure}

In Fig.~\ref{CaptureCaseTwo} capture probabilities are displayed for a case given by the parameters $D_V=\sigma_V=3$ and $l=1$. The complex energy spectrum is shown in the first panel. It is seen that the system supports 5 bound states and two resonance states between the first and second threshold. The next panels give capture probabilities for photon and phonon emission, respectively. For the former, this have also been shown for positive and negative parity separately. Two peaks are clearly seen -- one for each resonance. Moreover, it is seen that the resulting population of bound states reflects the parity of each of the two resonances. In this case, a resonance peak is seen for the phonon emission as well. Any peak near the resonance with position close to $-2$ is hard to see, however, for the same reason as above. The dispersion relation used here corresponds to longitudinal acoustic phonons in the $[111]$-direction for GaAs \cite{Ram1985}.

\section{Conclusions}
\label{conclusions}

Collision dynamics between two electrons in a quantum wire where one of them is initially trapped in a quantum dot have been investigated. This has been done within a theoretical framework in which the concept of absorbing boundaries have been generalized to apply to many-particle systems.
We have seen that, as expected, the collision dynamics is strongly influenced by the presence of doubly excited states. Specifically, it is seen that the reflection coefficient features a pronounced peak when the energy of the incident electron is such that it is resonant with an excitation to such doubly excited states. It was seen that this phenomena may lead to a enhanced probability of capture via spontaneous emission. Thus, by tuning the energy of the incident electron, doubly excited states may be exploited in order to facilitate the population of quantum dots.

\section{Acknowledgements}

This work has benefited from insightful advise and useful source code provided by dr. Simen Kvaal. Insightful advise have also been provided by dr. Simen Bræck, dr. Joakim Bergli and prof. Yurij Galperin. The author is also grateful for computer resources provided at the Department of Physics at Oslo University.



\end{document}